\documentstyle[gsirep]{article}
\newcommand\ba{\begin{equation}}
\newcommand\ea{\end{equation}}
\newcommand\be{\begin{eqnarray}}
\newcommand\ee{\end{eqnarray}}

\title{Nonperturbative Flow Equations at finite Temperature}
\author{B.-J. Schaefer and H.-J. Pirner\\
         Institut f\"ur Theor. Physik, University of Heidelberg}
\date{}
\begin{document}
\maketitle

\noindent 
QCD incorporates different relevant excitations at different length
scales. At short distances the relevant degrees of freedom are
quarks and gluons while particles which are observed at large scales
are colorless mesons and hadrons. The theoretical challenge is to
find a bridge between one set of degrees of freedom (quarks and
gluons) to another (mesons and hadrons). As a guiding principle
chiral symmetry, which is broken in the vacuum is used. It is 
believed that the linear $\sigma$-model is an effective description 
for QCD for scales below the mesonic compositeness scale. It
also shares an adequate realism with a good practicality in 
applications. 

Recently, renormalization group methods have been applied to
calculate the parameters of the $\sigma$-model for all resolution
scales. The fundamental idea is to follow the dynamics of the 
system by integrating out quantum fluctuations in infinitesimal
intervals from a high momentum scale to the far infrared. In 
practice these so-called exact renormalization group  flow
equations have to be truncated in order to obtain a finite
dimensional set of differential equations which can be solved
numerically. They allow to treat the critical fluctuations of the
long range $\sigma$-field near the second order chiral phase
transition at finite temperature for two massless quark flavors, 
where mean field methods or sub summations of the effective
mass type are insufficient.

In our work \cite {bjhj} we follow the spirit of this renormalization group 
approach and parametrize the shape of the effective mesonic 
potential of the linear $\sigma$-model, where we only allow
quartic and quadratic couplings. This approximation together
with a novel heat-kernel infrared cutoff prescription allows
us to derive new and very transparent formulae for the 
evolution equations. Our cutoff function is different from 
previous work \cite {jw} and it allows us to calculate analytically the
threshold functions, which arise in the flow equations and describe
the smooth decoupling of massive modes from the evolution 
towards small infrared scales. This work enables us to get analytical
insight into the physics inherent in the method of evolution
equations.

For high resolution a typical scale
associated with perturbative QCD is $\Lambda \geq 1.5$ GeV.
RHIC and LHC physics for particles with $p_{\perp} \geq \Lambda$
will be dominated by such processes. 
Since the strength of the QCD interactions increases with 
decreasing momentum transfer, characteristic $\bar q q$ bound 
states will form and influence the dynamics
at larger distances. A typical resolution where these processes start
to become important
is $\Lambda_{\chi SB} \approx 1.0$ GeV.
Below this scale chiral symmetry is spontaneously broken and
the vaccum acquires a nonvanishing quark and/or meson condensate.
We think that our heat-kernel cutoff corresponds to the resolution
scale of the photon $Q^2$ in deep inelastic scattering, where around
$Q^2 \approx 1$ GeV$^2$ the behavior of the structure function 
$F_2$ changes qualitatively indicating the phenomenon of chiral 
symmetry restoration. We use this experimental hint for the
transition around 1 GeV as input to our calculation.

In the broken phase the constituent quark mass will  increase 
with decreasing resolution and remain finite down to 
$\Lambda \approx \Lambda_{QCD}$. In the interval 
$\Lambda_{QCD} \leq k \leq \Lambda_{\chi SB}$ the dynamics is
governed by constituent quarks interacting via pions and $\sigma$-mesons.
One sees nicely in our approach
how the different quantum fluctuations from
the $\sigma$-mesons and constituent 
quarks become unimportant when the infrared scale 
parameter becomes smaller than the respective masses
of these states. These modes then decouple
from the further evolution, leaving the zero mass pions alone
in the evolution.

One also recognizes the different signs of the bosonic and fermionic
contributions to the flow equations. The bosons lead to an infrared
stable (ultraviolet unstable) coupling, whereas the fermions counteract this
tendency. Going from high scales $k$ to low $k$ one sees that the 
mesonic self-interaction $\lambda_k$ balances at intermediate values
of $k$, whereas in the far infrared the boson term wins.
At $k=k_{\chi SB} \approx 800$ MeV 
the $\beta$-function of
$\lambda_k$ is continuous and the mass parameter $m_k$ of the 
mesonic potential tends to zero signaling symmetry breaking. 
The vacuum expectation value $\phi_k$ stabilizes at small
values of $k$ and the evolution ends with 
$\lim_{k \to 0} \phi_k = f_\pi$.

The $k_{\chi SB}$ scale found in this calculation is somewhat smaller
than the resolution $Q$ of the photon in deep inelastic scattering,
but this result could be improved by a more sophisticated calculation 
with a running Yukawa coupling and wave function renormalizations.

After  having solved the evolution equations at zero temperature for
reasonable starting values of the coupling constants,
we  pursue the evolution at finite 
temperature.
Here the relevant parameter is the
ratio of the temperature over the infrared scale
parameter. 
Decoupling now sets in when the ratio of masses plus the
Matsubara frequencies over the
infrared scale becomes large.
At high temperatures the summation over Matsubara frequencies
is dominated by the lowest mode, thereby reducing softly the
dynamics to the corresponding three-dimensional field
theory, which is the purely bosonic $O(4)$-model.
We find in the chiral limit a second order phase transition at
a critical temperature $T_c \approx 130$ MeV.
Power law behavior is also visible in the order parameter leading
to the critical index $\beta=0.40$ which is in good
agreement with  results from Monte Carlo calculations. 

\end{document}